\documentclass[preprint,showpacs,preprintnumbers,amsmath,amssymb]{revtex4}
\usepackage{graphicx}
\usepackage{dcolumn}
\usepackage{bm}

\begin{document}

\title{Mesoscopic model for nano-channel confined DNA 
}

\author{Marco Zoli}

\address{School of Science and Technology \\  University of Camerino, I-62032 Camerino, Italy \\ marco.zoli@unicam.it}

\begin{abstract}
I present a Hamiltonian model and a computational method suitable to evaluate structural and thermodynamic properties of helical molecules embedded in crowded environments which may confine the space available to the base pair fluctuations. It is shown that, for the specific case of a short DNA fragment in a nanochannel, the molecule is markedly over-twisted and stretched by narrowing the width of the channel.
\end{abstract}


\maketitle

The stability properties of ds-DNA are important in a number of molecular biology techniques (e.g., polymerase chain reaction) and nanotechnological devices (e.g., DNA-based sensors) in which short synthetic  DNA's are used as a recognition element by virtue of the peculiar Watson-Crick base pairing, allowing for selective hybridization with a target sequence. The helix stability is also key to single molecule denaturation experiments, recently used in combination with nanochannel arrays, which confine and stretch the DNA molecule to be analyzed. 
The DNA properties are strongly affected in confined conditions as those which occur \textit{in vivo} in the crowded environments of cells where macro-molecules \textit{i)} reduce the free volume for base pair fluctuations thus suppressing the melting entropy, \textit{ii)} interfere with the dynamics of DNA looping thus affecting the speed of gene activation or repression.
While considerable amount of experimental work has been carried out over the last decades to investigate the relation among macro-molecular crowding, DNA dynamics and its biological functioning, much less theoretical studies have been produced so far on DNA in confining conditions. 
Here I focus on the interplay between DNA structure and environment analyzing how helical conformation and molecule size can be modified passing through a nanochannel of variable width.

\section{Model}

\begin{figure}
\includegraphics[height=7.0cm,width=7.5cm,angle=-90]{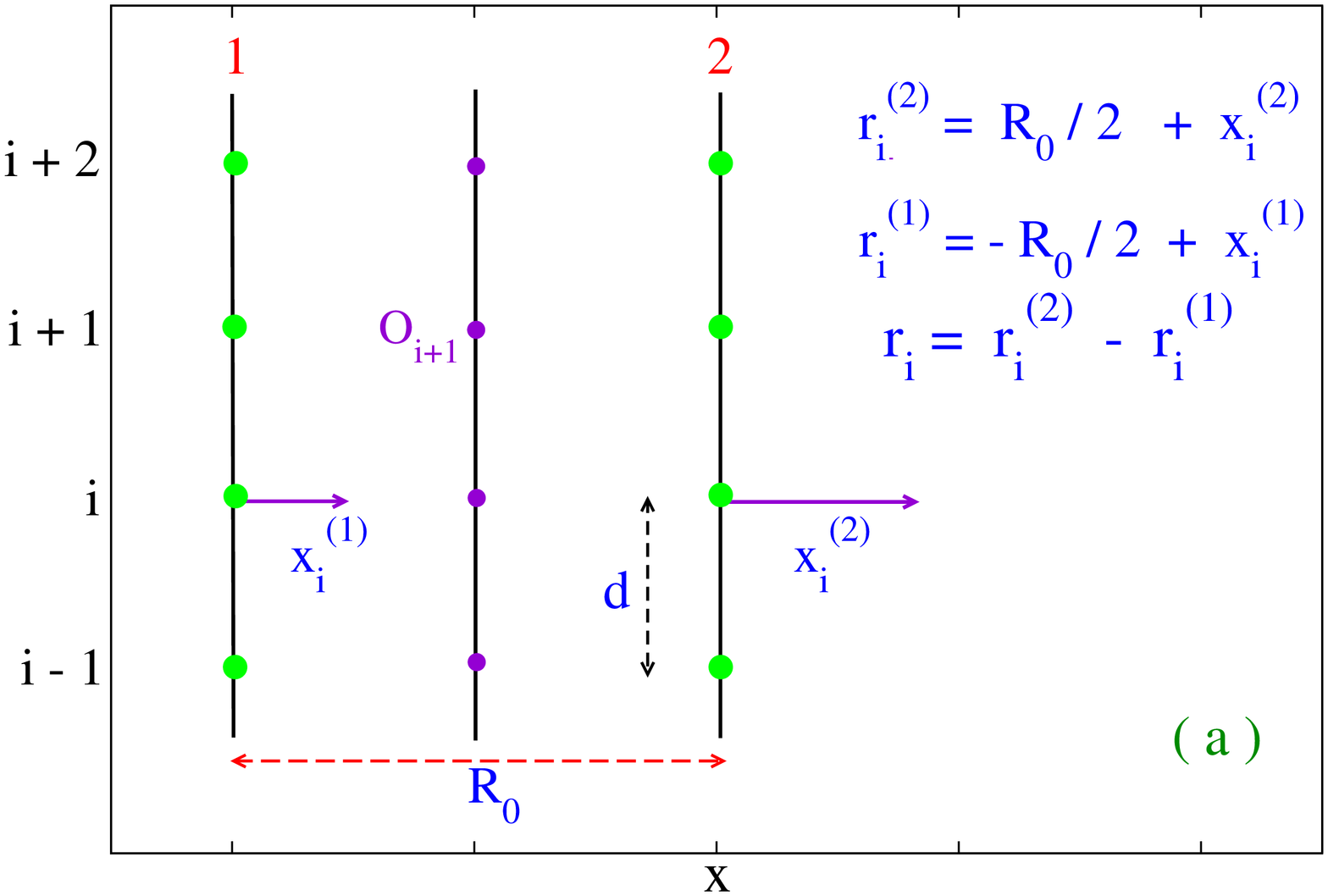}
\includegraphics[height=7.0cm,width=7.5cm,angle=-90]{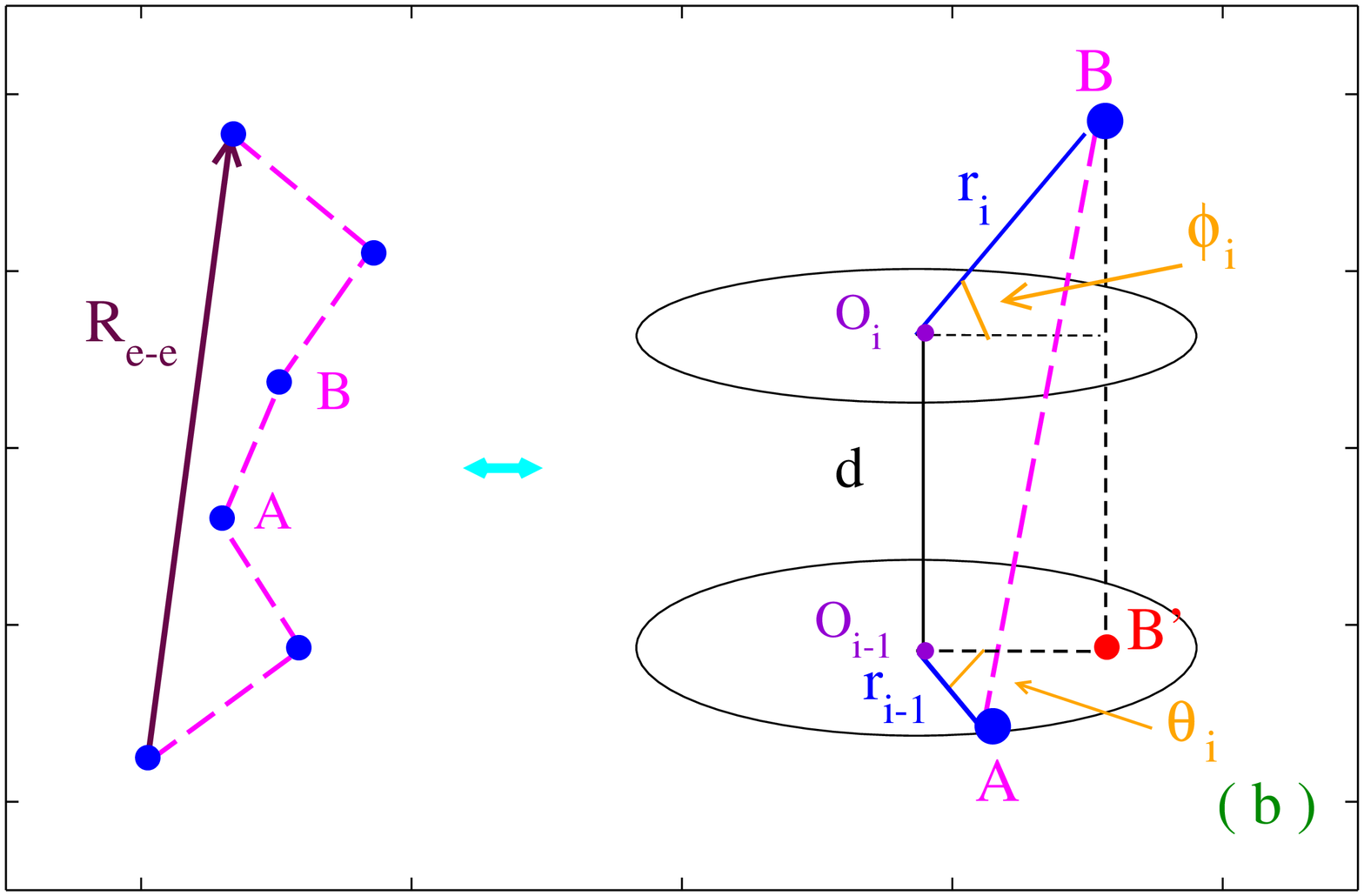}
\caption{\label{fig:1}(Color online) Schematic of (a) a ladder model and (b) a 3D model with radial and angular degrees of freedom for a helical molecule with point-like base pairs.
}
\end{figure}

The study is based on a coarse grained Hamiltonian model which describes the helical molecule at the level of the base pair (bp) \cite{io09,io10,io11a,io11b,io12,io13a,io13b,io15}. In a simple ladder model, the point-like bases are arranged at a fixed rise distance $d$ along the complementary strands (Fig.~\ref{fig:1}(a)). The two mates of the $i-th$ bp are assumed to fluctuate by $x_{i}^{(1,2)}$ around their equilibrium positions and the relative distance between the mates is denoted by $r_{i}$ which is measured with respect to the bare helix diameter $R_{0}$. While the ladder model offers a convenient description of the DNA molecule close to the melting transition, a more realistic mesoscopic representation of the double helix is attained by the 3D model in Fig.~\ref{fig:1}(b) which I have developed over the last years \cite{io18b} to account for the structural and thermodynamical properties of short double stranded fragments. Essentially the model assumes that adjacent bps along the molecule backbone, described respectively by the above defined radial distances $r_{i}, \, r_{i-1}$, can be twisted by the angle $\theta_i$ and bent by $\phi_i$ with respect to each other. This permits to introduce fluctuational effects in the model as the stacking distance 
between adjacent blue dots in Fig.~\ref{fig:1}(b), $\overline{AB} \equiv  \overline{d_{i,i-1}}$,  is a function both of the radial and angular variables.

The quantitative analysis of the model relies on a mesoscopic Hamiltonian which contains the main forces stabilizing the molecule i.e., \textit{1)} the hydrogen bonds between complementary bases, described by an effective one particle potential $V_{1}[r_i]$ with repulsive core and \textit{2)} the intra-strand stacking potential between adjacent bps along the molecule axis, described by a two particles potential $V_{2}[ r_i, r_{i-1}, \phi_i, \theta_i]$ \cite{io17}. The model potential parameters are taken in the usual range \cite{io16a,io16b} so as to fit available thermodynamic and elastic data. 

The equilibrium statistics of the system is obtained by a computational method based on the finite temperature path integral method \cite{fehi} whose application to DNA has been elucitated in several papers \cite{io14a,io14b,io14c}. 
Basically the base pair fluctuations $r_{i}$ are treated as imaginary time dependent paths
$r_i(\tau)$ with $\tau \in [0, \beta]$ and  $\beta=\,(k_B T)^{-1}$. $k_B$ is the Boltzmann constant and $T$ is the temperature. 
By virtue of the space-time mapping technique,  the partition function $Z_N$   is given by an integral over closed paths, $r_i(0)=\,r_i(\beta )$. As a consequence, the paths $r_i(\tau)$ can be expanded in a Fourier series, \, $r_i(\tau)=\, (r_0)_i + \sum_{m=1}^{\infty}\Bigl[(a_m)_i \cos( \frac{2 m \pi}{\beta} \tau ) + (b_m)_i \sin(\frac{2 m \pi}{\beta} \tau ) \Bigr]$, \, whose coefficients yield, for any base pair, a set of possible fluctuations statistically contributing to $Z_N$. The code includes in the computation only those combinations of Fourier coefficients yielding radial fluctuations which are physically consistent with the constraints imposed by the model potential. Explicitly, for the chain in  Fig.~\ref{fig:1}(b) made of $N$ bps of reduced mass $\mu$,  $Z_N$ reads:

\begin{eqnarray}
& &Z_N=\, \oint Dr_{1} \exp \bigl[- A_a[r_1] \bigr]   \prod_{i=2}^{N}  \int_{- \phi_{M} }^{\phi_{M} } d \phi_i \int_{- \theta_{M} }^{\theta _{M} } d \theta_{i} \oint Dr_{i}  \exp \bigl[- A_b [r_i, r_{i-1}, \phi_i, \theta_i] \bigr] \, , \nonumber
\\
& &A_a[r_1]= \,  \int_{0}^{\beta} d\tau \Bigl[ \frac{\mu}{2} \dot{r}_1(\tau)^2 + V_{1}[r_1(\tau)] \Bigr] \, , \nonumber
\\
& &A_b[r_i, r_{i-1}, \phi_i, \theta_i]= \,  \int_{0}^{\beta} d\tau \Bigl[ \frac{\mu}{2} \dot{r}_i(\tau)^2 + V_{1}[r_i(\tau)] + V_{2}[ r_i(\tau), r_{i-1}(\tau), \phi_i, \theta_i]  \Bigr]      \, . 
\label{eq:01}
\end{eqnarray}

$A_a[r_1]$ is treated separately as the first base pair has no preceding neighbour along the stack. $\phi_{M}=\,\pi /2$ and $\theta_{M}=\,\pi /4$ are the bending and twisting integration cutoffs, respectively. 

$\oint {D}r_i$ is the integration measure over the above defined Fourier coefficients:

\begin{eqnarray}
& &\oint {D}r_{i} \equiv {\frac{1}{\sqrt{2}\lambda_{cl}}} \int_{-\Lambda_{T}^{0}}^{\Lambda_{T}^{0}} d(r_0)_i \prod_{m=1}^{\infty}\Bigl( \frac{m \pi}{\lambda_{cl}} \Bigr)^2 \int_{-\Lambda_{T}}^{\Lambda_{T}} d(a_m)_i \int_{-\Lambda_{T}}^{\Lambda_{T}} d(b_m)_i \, , \, 
\label{eq:02}
\end{eqnarray}

with  $\lambda_{cl}$ being the classical thermal wavelength. The cutoffs \, $\Lambda_{T}^{0}$ and  $\Lambda_{T}$ are determined for computational purposes using a fundamental property of the path integration method, i.e., $\oint {D}r_i$  normalizes the kinetic action:

\begin{eqnarray}
\oint {D}r_i \exp\Bigl[- \int_0^\beta d\tau {\mu \over 2}\dot{r}_i(\tau)^2  \Bigr] = \,1 \, .
\label{eq:03} \,
\end{eqnarray}

Hence, using the path Fourier expansion, from Eqs.~(\ref{eq:02}),~(\ref{eq:03}), one gets: \, ${\Lambda^0_T}=\,\lambda_{cl}/\sqrt{2}$ \, and \, $\Lambda_T =\,{{U \lambda_{cl}} / {m \pi^{3/2}}}$ \,
with $U$ being a dimensionless parameter. Setting \, $U=\,12$, we attain numerical convergence in the computation of Eq.~(\ref{eq:01}) and include large amplitude base pair fluctuations.

In general, the outlined computational method proves suitable to describe the effects on a single DNA molecule brought about by a crowders distribution which is expected to confine the phase space available to the base pair fluctuations. Such effects can be simulated by replacing $U$ by a site dependent cutoff, $\, U \rightarrow U - C(i)$, where  $C(i)$ is the function accounting for the confinement caused by the specific crowders surrounding the helical fragment.
In the case of a nano-channel, the DNA molecule passing through the device is uniformly confined, i.e., for all base pairs in the chain, the fluctuations should shrink by the same amount. Then, we incorporate such conditions within our model by assuming a site independent  \, $C(i)=\,\gamma U$,
where $\gamma \in (0,1)$ is a tunable parameter whose strength is inversely proportional to the channel width.

\section{Results}

The model is applied to compute the DNA size and shape in thermal equilibrium with the confining environment. The statistical mechanics of the system is worked out by performing the ensemble averages for the molecule physical properties via Eqs.~(\ref{eq:01}),~(\ref{eq:02}) for any choice of $\gamma$.
While the model generally assumes an helical molecular shape,  the specific twist conformation of the fragment for the state of thermodynamic equilibrium is \textit{a priori} unknown. Then, the idea inspiring our code is that of taking the number of base pairs per helix turn (i.e., the helical repeat $h$) as an input value and, for such value, calculating the free energy $\, F=\, -\beta ^{-1} \ln Z_N$. 

More precisely, the code samples a broad range of J-values for $h$  around the experimental $\,h^{exp}=\,10.4$ value, albeit estimated for kilo-base long DNA, and
for the $j-th$ value in such range (j=\,1,...,J),  a iterative procedure  yields the ensemble averaged twist angles $<\theta_{i}>$ for all base pairs in the chain.
Once $< \theta_N >$ is obtained, one gets the $j-th$ helical repeat averaged over the statistical ensemble of Eq.~(\ref{eq:01}): $ < h >_{j}=\,{2\pi N}/{< \theta_N >}$.
As the procedure is repeated for any initial $h_j$, one derives a set of $J$ average twist conformations. Among the latter, the value ($< h >_{j^{*}}$)  that corresponds to the state of thermodynamic equilibrium is finally selected by minimizing $F$.
For any twist conformation defined by $< h >_{j}$, the code can also deliver the macroscopic average parameters providing a measure for the global DNA size \cite{io18a}. In particular, we focus hereafter on the end-to-end distance depicted in Fig.~\ref{fig:1}(b) and calculated its average as:
$< R_{e-e} > =\, \biggl < \biggl| \sum_{i=2}^{N}  \overline{d_{i,i-1}} \biggr| \biggr > $.

\begin{figure}
\includegraphics[height=14.0cm,width=7.0cm,angle=-90]{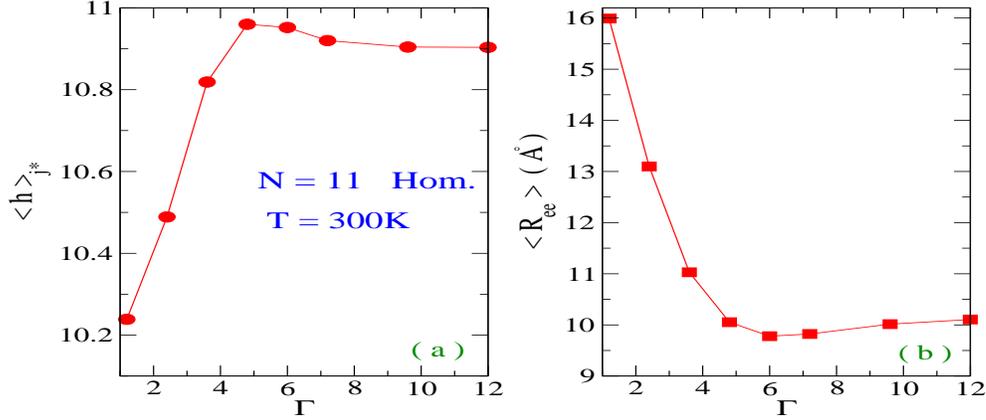}
\caption{\label{fig:2}(Color online) Room temperature (a) average helical repeat and (b) average end to end distance of a short fragment in a nano-channel versus the confinement strength. 
}
\end{figure}

The results for $< h >_{j^{*}}$ and the associated $< R_{e-e} >$  are displayed in Fig.~\ref{fig:2} as a function of $\Gamma  \equiv \, U (1 - \gamma)$. The simulation is for a short homogeneous fragment with $N=\,11$, at room temperature. It is found that the molecule is substantially stable (or even slightly untwisted for small $\gamma$'s) whereas, once the confinement becomes sizable ($\Gamma < 5$), the helix markedly over-twists and consistently shows a significant increase of $< R_{e-e} >$. The latter is elongated to values even larger than $15 \AA$ in the regime in which all base pairs are strongly confined.
Then, the model and the computational method can quantitatively estimate the stretching of a DNA chain in a nano-channel and also account for the peculiar interplay between helix overtwisting and stretching observed in the different context of force-extension experiments on single molecules \cite{io18c}.  The method can be further extended to deal with crowders distributions operating non uniform confinements of heterogeneous fragments with larger $N$ at various $T$ \cite{io19}.

\end{document}